\documentclass[unsortedaddress,preprint,show pacs]{revtex4-1}
\usepackage{graphicx}
\usepackage{dcolumn}
\usepackage{amsmath}
\usepackage{amssymb}
\usepackage{amsfonts}
\usepackage{bm}
\usepackage{color}
\newcommand{\be}{\begin{equation}}
\newcommand{\ee}{\end{equation}}
\newcommand{\ba}{\begin{eqnarray}}
\newcommand{\ea}{\end{eqnarray}}

\begin{document}
\title{Shapes of a liquid droplet in a periodic box}
\author{Santi Prestipino$^{1,2}$\footnote{Corresponding author. Email: {\tt sprestipino@unime.it}}, Carlo Caccamo$^1$\footnote{Email: {\tt caccamo@unime.it}}, Dino Costa$^1$\footnote{Email: {\tt dcosta@unime.it}}, Gianpietro Malescio$^1$\footnote{Email: {\tt malescio@unime.it}}, and Gianmarco Muna\`o$^1$\footnote{Email: {\tt gmunao@unime.it}}}
\affiliation{$^1$Universit\`a degli Studi di Messina, Dipartimento di Fisica e di Scienze della Terra, 
Contrada Papardo, I-98166 Messina, Italy\\$^2$CNR-IPCF, Viale F. Stagno d'Alcontres 37, I-98158 Messina, Italy}
\date{\today}
\begin{abstract}
Within the coexistence region between liquid and vapor the equilibrium pressure of a simulated fluid exhibits characteristic jumps and plateaus when plotted as a function of density at constant temperature. These features exclusively pertain to a finite-size sample in a periodic box, as they are washed out in the bulk limit. Below the critical density, at each pressure jump the shape of the liquid drop undergoes a morphological transition, changing from spherical to cylindrical to slab-like as the density is increased. We formulate a simple theory of these shape transitions, which is adapted from a calculation originally developed by Binder and coworkers [{\em J. Chem. Phys.} {\bf 120}, 5293 (2004)]. Our focus is on the pressure equation of state (rather than on the chemical potential, as in the original work) and includes an extension to elongated boxes. Predictions based on this theory well agree with extensive Monte Carlo data for the cut-and-shifted Lennard-Jones fluid. We further discuss on the thermodynamic stability of liquid drops with shapes other than the three mentioned above, like those found deep inside the liquid-vapor region in simulations starting from scratch. Our theory classifies these more elaborate shapes as metastable.
\end{abstract}
\maketitle

\section{Introduction}

Statistical mechanics gives a general prescription for extracting equilibrium properties from a system Hamiltonian. To this aim, the system partition function should be determined in any of a number of equivalent descriptions (ensembles), a task requiring the evaluation of a complicate multidimensional integral. In practice, this can only be accomplished in a few (mainly trivial) cases, which thus obliges one to go over to partial calculations and approximate theories, often of limited scope. With the advent of computer simulation, the analytically intractable program of statistical mechanics could eventually be attacked and its solution finally came within reach, at least for finite, not too large systems.

Occasionally, however, numerical simulation may lead to misconceptions. An example is the loop found in the pressure equation of state constructed by simulation within the coexistence region of liquid and vapor. Similar loops are found in the van der Waals theory of condensation, which uses the double-tangent construction to make the chemical potential everywhere concave as a function of pressure at constant temperature. As a result, the so-called metastable and unstable branches of the original equation of state are thrown out as unphysical. It is so customary to associate condensation with van der Waals theory, that it would tempting to interpret the loops canonical-ensemble simulations always produce in isotherms of intensive variables as van der Waals loops and, therefore, to read them as a sign of the entrance of vapor in a metastable regime. However, as originally remarked in Refs.\,\cite{Mayer,Binder1}, the non-concave regions observed in the equation of state of a finite system are fully equilibrium features arising from the use of periodic boundary conditions in the simulation. In particular, the first inversion of concavity encountered corresponds to the first occurrence of liquid-vapor separation. We have recently showed that, despite their fake character, one can take advantage of these pressure loops to extract, by plain thermodynamic integration, the right liquid-vapor coexistence parameters~\cite{Abramo1,Abramo2}.

In a series of papers~\cite{Binder1,MacDowell1,MacDowell2,Binder2}, Binder and coworkers carried out extensive grand-canonical Monte Carlo (MC) simulations of the Lennard-Jones (LJ) fluid in a periodic cubic box, biasing the sampling in such a way as to constrain the system to stay in the two-phase region. They found a whole sequence of so-called {\em shape transitions} between various ``phases'' differing in the shape of the liquid droplet coexisting with vapor. Specifically, upon increasing the system density $\rho$ the liquid drop changed from spherical (``sph'') to cylindrical (``cyl'') to slab-like (``slab''). Upon increasing $\rho$ further, the reversed sequence of transitions was observed, with interchanged roles between liquid and vapor. At each rearrangement of the liquid-vapor interface, the chemical potential undergoes a sharp drop, followed by a density interval where it stays nearly constant (this region will be referred to in the following as a ``plateau''). MacDowell {\it et al.} have successfully analysed these shape transitions by means of a capillary-drop theory~\cite{MacDowell1,MacDowell2}. In a further series of paper~\cite{Schrader,Block,Troester}, Binder and his group exploited their extremely accurate simulation results to extract information about the interface free energy of curved liquid-vapor interfaces, and thus have access to the Tolman length~\cite{Tolman}, an all-useful parameter in nucleation theory~\cite{Kashchiev,Koga,Prestipino1} (the only caution here would be that in Refs.\,\cite{Schrader,Block,Troester} the interfaces were taken at full equilibrium rather than under the metastable conditions typical of nucleation experiments).

We hereby consider the phenomenon of condensation in canonical-ensemble simulations in the light of a yet different theory which, at variance of van der Waals theory, right from the outset takes into account the periodic repetition of the system in space. The present theory originates from the few-line calculation presented in Sect.\,II-A of Ref.\,\cite{MacDowell1}, but now putting the emphasis on the pressure -- rather than the chemical potential -- equation of state, since it is the pressure that is directly accessible in a canonical-ensemble simulation. The theory is further refined by an extension to elongated-cubic simulation boxes. As a matter of principle, geometric shapes other than spherical, cylindrical, and tetragonal may also occur for the liquid drop in equilibrium with vapor, and this possibility will be looked closely within our theoretical framework.

The remainder of the paper is structured as follows. In Sec. II, we expose the details of our theory. In Sec. III, we compare simulation results for the cut-and-shifted Lennard-Jones potential with theoretical predictions, discussing the relative stability of spherical, cylindrical, and slab-like drops as a function of the box aspect ratio. The stability of other interface shapes, like those observed in simulation runs performed independently near the cylinder-slab transition density~\cite{Abramo2}, is also studied. Finally, we give our conclusions in Sec. IV.

\section{Theory}
\setcounter{equation}{0}
\renewcommand{\theequation}{2.\arabic{equation}}

In order to determine the pressure behavior of a fluid near above the vapor coexistence density, the idea originally put forward in Ref.~\cite{MacDowell1} is to compare the Helmholtz free energy $F$ of the homogeneous vapor with that of various competing heterogeneous ``phases'', differing in the shape of the liquid drop in thermal equilibrium with vapor. Since the thermodynamically stable phase for fixed temperature $T$, volume $V$, and particle number $N$ is the one with minimum $F$, the drop adopts the shape which makes the total free energy as small as possible, consistent with the amount of liquid present at the given density $\rho=N/V$.

If the focus is on the pressure equation of state, rather than on the chemical potential, it is convenient to express the free energy in terms of the specific volume $v=1/\rho$. This choice offers a number of practical advantages as it will be made clear below. Let us first estimate the free energy of the {\em homogeneous vapor} as a function of $v$ at fixed $T$. Denoting $f(T,v)=F/N$ the free energy per particle, a general relation valid up to second-order terms in the deviations $\Delta T=T-T_0$ and $\Delta v=v-v_0$ from a given state point ${\cal S}_0=(T_0,v_0)$ is:
\be
f=f_0-\frac{S}{N}\Delta T-P\Delta v-\frac{c_V}{2T}\Delta T^2-\frac{\alpha_P}{K_T}\Delta v\Delta T+\frac{1}{2vK_T}\Delta v^2\,,
\label{2-1}
\ee
with $f_0=f(T_0,v_0)$. In the latter equation $S$ is the entropy, $c_V$ is the constant-volume specific heat, $\alpha_P$ is the isobaric expansion coefficient, and $K_T$ is the isothermal compressibility, all computed at ${\cal S}_0$. Choosing ${\cal S}_0$ to be the condensation point at temperature $T$, for $\Delta T=0$ and $\Delta v=v-v_{\rm v}$ ($v_{\rm v}\equiv 1/\rho_{\rm v}$ being the vapor specific volume at coexistence) it follows from Eq.\,(\ref{2-1}) that
\be
\Delta F_{\rm hom}\equiv F-F_{\rm v}=P_{\rm v}(v_{\rm v}-v)N+\frac{1}{2v_{\rm v}K_{\rm v}}(v-v_{\rm v})^2N\,,
\label{2-2}
\ee
denoting $P_{\rm v}$ and $K_{\rm v}$ the condensation pressure and isothermal compressibility of the bulk vapor at $\rho_{\rm v}$ (in the following, we use ``v'' and ``l'' subscripts to denote bulk properties of the coexisting vapor and liquid).

For a vapor system in equilibrium with a {\em spherical} drop of liquid hosting $N_{\rm l}$ atoms, in the capillarity approximation the cost of droplet formation is
\ba
\Delta F_{\rm sph}\equiv F_{{\rm v}+{\rm l}}-F_{\rm v}&=&(N-N_{\rm l})f_{\rm v}+N_{\rm l}f_{\rm l}+(36\pi)^{1/3}\gamma(N_{\rm l}v_{\rm l})^{2/3}-Nf_{\rm v}
\nonumber \\
&=&N_{\rm l}(f_{\rm l}-f_{\rm v})+(36\pi)^{1/3}\gamma(N_{\rm l}v_{\rm l})^{2/3}\,,
\label{2-3}
\ea
where $\gamma$ is the surface tension at temperature $T$ ({\it i.e.}, the free energy of the planar liquid-vapor interface). In deriving Eq.\,(\ref{2-3}), curvature corrections to surface tension~\cite{Prestipino1} as well as anisotropy effects~\cite{Prestipino2,Prestipino3} have been neglected. Assuming for the vapor and liquid fractions the same chemical potential and pressure as in the bulk limit, the difference $f_{\rm l}-f_{\rm v}$ becomes $P_{\rm v}(v_{\rm v}-v_{\rm l})$. Moreover $Nv=N_{\rm l}v_{\rm l}+(N-N_{\rm l})v_{\rm v}$, as no volume is attached to the interface (this is to be contrasted with the lever-rule estimate of $V_{\rm l}$ made in Ref.\,\cite{MacDowell1}, which implicitly assumed zero adsorption for the interface). In conclusion, we get:
\be
\Delta F_{\rm sph}=P_{\rm v}(v_{\rm v}-v)N+(36\pi)^{1/3}\gamma v_{\rm l}^{2/3}\left(\frac{v_{\rm v}-v}{v_{\rm v}-v_{\rm l}}\right)^{2/3}N^{2/3}\,.
\label{2-4}
\ee
From Eqs.\,(\ref{2-2}) and (\ref{2-4}), we readily derive the pressures of the two ``phases'':
\be
P_{\rm hom}=P_{\rm v}+\frac{\rho-\rho_{\rm v}}{K_{\rm v}\rho}\,\,\,\,\,\,{\rm and}\,\,\,\,\,\,P_{\rm sph}=P_{\rm v}+\frac{2}{3}(36\pi)^{1/3}\frac{\gamma\rho_{\rm v}}{\rho_{\rm l}-\rho_{\rm v}}\left(\frac{\rho(\rho_{\rm l}-\rho_{\rm v})}{\rho-\rho_{\rm v}}\right)^{1/3}N^{-1/3}\,.
\label{2-5}
\ee
At variance with Ref.\,\cite{MacDowell1} $P_{\rm hom}$ is a non-linear function of $\rho$, which makes it better suited to reproduce the true pressure behavior close to $\rho_{\rm v}$ (see Fig.\,1 below). By the way, the $(\rho-\rho_{\rm v})^2$ term in $P_{\rm hom}$ may not be the exact one since Eq.\,(\ref{2-2}) is only a second-order truncated expansion.

When the liquid drop is a {\em cylinder} extending along the shorter edge of a cuboidal box, hence of length $L_x=L_y=L_z/a=(V/a)^{1/3}$ (where $a>1$ is the box aspect ratio), the free-energy excess over the vapor at $\rho_{\rm v}$ becomes:
\be
\Delta F_{\rm cyl}=N_{\rm l}(f_{\rm l}-f_{\rm v})+2\pi\gamma r_{\rm cyl}L_x\,,
\label{2-6}
\ee
where the radius $r_{\rm cyl}$ follows from
\be
\pi r_{\rm cyl}^2L_x=N_{\rm l}v_{\rm l}\,\,\,\,\,\,{\rm with}\,\,\,\,\,\,N_{\rm l}=\frac{v_{\rm v}-v}{v_{\rm v}-v_{\rm l}}N\,.
\label{2-7}
\ee
Upon inserting Eq.\,(\ref{2-7}) into (\ref{2-6}), we obtain:
\be
\Delta F_{\rm cyl}=P_{\rm v}(v_{\rm v}-v)N+2\pi^{1/2}\gamma a^{-1/6}\rho^{-2/3}\left(\frac{\rho-\rho_{\rm v}}{\rho_{\rm l}-\rho_{\rm v}}\right)^{1/2}N^{2/3}\,,
\label{2-8}
\ee
thus yielding
\be
P_{\rm cyl}=P_{\rm v}-\frac{\pi^{1/2}\gamma v_{\rm l}^{1/2}}{3a^{1/6}v^{5/6}}\frac{v_{\rm v}-4v}{v_{\rm v}-v_{\rm l}}\left(\frac{v_{\rm v}-v_{\rm l}}{v_{\rm v}-v}\right)^{1/2}N^{-1/3}\,.
\label{2-9}
\ee

Finally, when the liquid fills a {\em slab} lying perpendicularly to the longer box edge ($L_z$), the free-energy excess becomes:
\be
\Delta F_{\rm slab}=N_{\rm l}(f_{\rm l}-f_{\rm v})+2\gamma L_x^2=P_{\rm v}(v_{\rm v}-v)N+\frac{2\gamma}{a^{2/3}}v^{2/3}N^{2/3}\,,
\label{2-10}
\ee
whence a pressure of
\be
P_{\rm slab}=P_{\rm v}-\frac{4\gamma}{3a^{2/3}}\rho^{1/3}N^{-1/3}\,.
\label{2-11}
\ee

In the present analysis, a crossing of free energies as a function of $\rho$ entails a change in the relative stability of two drop shapes. Far from being a first-order transition, which can only occur in the thermodynamic limit and is not accompanied by any pressure jump, this shape transition is an equilibrium finite-size effect promoted by periodic boundary conditions. In fact, all shape transitions become rounded crossovers when thermal fluctuations are taken into account (see Refs.\,\cite{MacDowell1,MacDowell2}). Upon equating the free energies two at a time, and providing that the sequence of ``phases'' is the same as found in Ref.\,\cite{MacDowell2}, we obtain the following formulae for the ``transition'' densities:
\ba
\rho_{{\rm hom}-{\rm sph}}&=&\rho_{\rm v}\left(1-(36\pi)^{1/4}\frac{(2K_{\rm v}\rho_{\rm v}\gamma)^{3/4}}{(\rho_{\rm l}-\rho_{\rm v})^{1/2}}N^{-1/4}\right)^{-1}\,;
\nonumber \\
\rho_{{\rm sph}-{\rm cyl}}&=&\rho_{\rm v}+\frac{4\pi}{81a}(\rho_{\rm l}-\rho_{\rm v})\,;
\nonumber \\
\rho_{{\rm cyl}-{\rm slab}}&=&\rho_{\rm v}+\frac{1}{\pi a}(\rho_{\rm l}-\rho_{\rm v})\,.
\label{2-12}
\ea
In particular, we note that the density range beyond $\rho_{\rm v}$ where the homogeneous vapor is {\em thoroughly stable} vanishes in the thermodynamic limit as $N^{-1/4}$. This $\rho$ interval should not be confused with the vapor metastability region, which instead is a kinetic concept more appropriate to bulk systems. The metastability region of vapor, which extends past the $T$-$P$ coexistence locus inside the liquid region, comprises all nominally-liquid states where a quenched vapor system can be maintained gaseous for a long time (that is, much longer than the typical observation times) before nucleation of the liquid occurs.

With a further little effort, one may also derive the size of the liquid droplet in its various conformations, obtaining:
\ba
r_{\rm sph}&=&\left(\frac{3}{4\pi}\frac{\rho-\rho_{\rm v}}{\rho_{\rm l}-\rho_{\rm v}}\right)^{1/3}\left(\frac{N}{\rho}\right)^{1/3}\,;
\nonumber \\
r_{\rm cyl}&=&\left(\frac{a^{1/3}}{\pi}\frac{\rho-\rho_{\rm v}}{\rho_{\rm l}-\rho_{\rm v}}\right)^{1/2}\left(\frac{N}{\rho}\right)^{1/3}\,;
\nonumber \\
d_{\rm slab}&=&a^{2/3}\frac{\rho-\rho_{\rm v}}{\rho_{\rm l}-\rho_{\rm v}}\left(\frac{N}{\rho}\right)^{1/3}\,,
\label{2-13}
\ea
where the latter quantity represents the thickness of the liquid slab.

As a final comment, we underline that the above theory would also apply with no modifications to the solid-liquid transition~\cite{Statt}. However, a testing of the theory against simulation data would actually be impossible in this case, because of the tendency of both phases to go deeply metastable (see, {\it e.g.}, Ref.~\cite{Abramo2}), a fact that generally prevents one from observing any shape transition. A notable exception is the numerical evidence reported in Fig.\,2 of Ref.\,\cite{Bernard}, where, thanks to a smart simulation method, the low system dimensionality, and very long runs, it was possible to construct a pressure equation of state showing a plateau in the density range corresponding to the slab formation.

\section{Assessment of the theory: Lennard-Jones model}
\setcounter{equation}{0}
\renewcommand{\theequation}{3.\arabic{equation}}

As originally shown in Ref.\,\cite{MacDowell2}, the sequence of shape transitions in a large periodic cubic box is expected to be hom $\rightarrow$ sph $\rightarrow$ cyl $\rightarrow$ slab, then followed by the reversed transition sequence with the role of vapor and liquid interchanged. This finding is confirmed by the results of our simulations for the cut-and-shifted LJ model and gratifyingly reproduced by our theory for $a=1$ (see Sect.\,III-A below). The possibility of other stable shapes, which as far as we know was never examined before, is discussed in the next Sect.\,III-B.

\subsection{Results from sequential simulations}

In a recent paper~\cite{Abramo2}, we performed extensive Metropolis MC simulations of the {\em cut-and-shifted LJ model} in the $NVT$ ensemble. This model is characterized by the following interaction potential:
\ba
u(r)=\left\{
\begin{array}{rl}
4\epsilon\left[(\sigma/r)^{12}-(\sigma/r)^6\right]-c\,, & \,\,\,{\rm for}\,\,r<r_{\rm cut}\\
0\,, & \,\,\,{\rm for}\,\,r>r_{\rm cut}
\end{array}
\right.
\label{3-1}
\ea
with $r_{\rm cut}=2.5\sigma$, where the constant $c$ is chosen in such a way as to make $u(r)$ everywhere continuous (from here onward, all quantities will be expressed in the units set by $k_{\rm B},\epsilon$, and $\sigma$, where $k_{\rm B}$ is Boltzmann's constant). The critical temperature of this system is slightly less than 1.10~\cite{Smit}. In particular, in Ref.\,\cite{Abramo2} we considered the behavior of the system along the isotherm $T=0.90$, to see whether in the liquid-vapor region equilibrium can be reached by traditional simulation methods with local moves. To this end we performed simulation runs in a sequence, starting at each density from the last system configuration produced in the previous run at a slightly smaller density. We thus showed that, anywhere within the coexistence region, heterogeneous equilibrium can be established in an affordable time. This is proved by the fact that we were able to obtain the known liquid-vapor coexistence densities by integrating the pressure equation of state across the two-phase region. At least this was the case for a system of $N=1372$ particles or smaller, while a larger system of 4000 particles was found to be plagued by metastability ({\it i.e.}, the data collected along the forward and backward trajectories were not the same, see Fig.\,1 below).

Having accurate simulation results available in the liquid-vapor region gives the opportunity to make a critical appraisal of the theory presented in Sect.\,II. For $T=0.90$, the liquid-vapor coexistence pressure and densities are $P_{\rm v}=0.03146,\rho_{\rm v}=0.0451$, and $\rho_{\rm l}=0.6649$, respectively~\cite{Abramo2}. In order to compute $K_{\rm v}$, we carried out a long $NPT$ simulation of the vapor at $P=P_{\rm v}$, eventually finding an isothermal compressibility of 45.04 (we computed $K_{\rm v}$ from the formula $K_T=\beta(\langle V^2\rangle-\langle V\rangle^2)/\langle V\rangle$, where $\beta=(k_{\rm B}T)^{-1}$ and $\langle\cdots\rangle$ is the isothermal-isobaric average). The only theory parameter left to set is $\gamma$, and we decided to choose it in such a way that $\rho_{{\rm hom}-{\rm sph}}$ roughly coincides with the location of the first pressure drop in the MC data for $N=4000$~\cite{Abramo2} (we thus obtained $\beta\gamma=0.19$). We report in Fig.\,1 the results from theory, and compare them against MC data. In the top panel, we show $\Delta F/N-P_{\rm v}(v_{\rm v}-v)$ for the homogeneous vapor (cf. Eq.\,(\ref{2-2})) and the various heterogeneous phases (cf. Eqs.\,(\ref{2-4}), (\ref{2-8}), and (\ref{2-10})). The sequence of stable conformations as a function of density is as expected, that is hom $\rightarrow$ sph $\rightarrow$ cyl $\rightarrow$ slab. In the middle panel of Fig.\,1 we compare the theoretical pressure (red line) with MC data from both forward- and backward-travelled paths~\cite{Abramo2}: we see an overall good agreement, especially in the slope of the pressure plateaus, but also the location of shape transitions would be well reproduced by the theory considering that the data from the backward trajectory are probably the closest to equilibrium (indeed, it is much easier for the system to preserve its structure during overcompression rather than when reducing the density below the transition threshold). Finally, the bottom panel in Fig.\,1 reports the characteristic size of the liquid drop in each ``phase''.

For high enough densities, other conformations of drop become available to the system, as seen in the complete MC pressure equation of state (Fig.\,2) where each inflection point marks a different shape transition. The theory for these further transitions can be formulated following the same steps as before, by everywhere exchanging vapor with liquid, and the results are shown in Fig.\,3 for the same parameters used to draw Fig.\,1. From a glance at Fig.\,3 we see that the accuracy of the theory is now poorer, though it would be safe to say that it still qualitatively reproduces the MC data. A worse agreement with Monte Carlo should anyway be expected just for the reason that equilibration is more difficult at high density and the spatial definition of the interface between liquid and vapor is also poorer. In summary, we reached a good agreement between simulation and theory by adjusting one single parameter.

In Ref.\,\cite{Abramo2} we also reported simulation data obtained from sequential runs of $N=1500$ Lennard-Jones particles in a periodic cuboidal box with edges in the ratio of 1:1:3, for $T=0.90$. These results revealed that the spherical ``phase'' is apparently never stable whereas the cylindrical ``phase'' is strongly reduced in extent. We tested this conclusion by the theory of Sect.\,II and we indeed found that, under compression, the homogeneous vapor first gives way to a cylindrical drop of liquid, which then changes to a slab upon increasing the density further (see Fig.\,4). Again, we see a clear correlation between the location of the inflection points in the pressure data and the transition thresholds derived from theory. As a last note we observe that, according to the same theory, the stability of the cylindrical drop progressively reduces upon increasing $a$, until the cylindrical phase disappears altogether (for $a\approx 10$) and the homogeneous vapor is thereafter directly transformed into the slab phase. The use of an elongated box is instrumental to obtaining a wider density interval for the study of the planar liquid-vapor interface, which could be useful for example in the determination of the surface tension as a function of temperature.

\subsection{Other drop shapes}

Our next point concerns the evidence, originally found in LJ-model simulations for $T=0.75$~\cite{Abramo2}, of an additional pressure ``plateau'' (in fact, a slightly inclined flat region) lying between the ``cylinder'' and the ``slab'' plateaus. The trick to obtain this result was to start the simulation from scratch at every density. We commented in Ref.\,\cite{Abramo2} that, in the interval of densities corresponding to the extra plateau, the liquid drop has the shape of a slab with a hole. 

We carried out $NVT$ MC simulations of the cut-and-shifted LJ fluid for $T=0.72$ in a periodic cubic box, for a number of densities close to $\rho_{\rm cyl-slab}\approx 0.25$ (with the values of $\rho_{\rm v}$ and $\rho_{\rm l}$ taken from Ref.\,\cite{Trokhymchuk}), always assuming a face-centered-cubic structure for the initial configuration of the run. Clearly, this may not be the best choice to study heterogeneous equilibria by simulation, since relaxation times would be much longer than those encountered when performing simulation runs sequentially. However, the point is that by this method we may find out long-lived metastable states that would not be reached by sequential simulations. For each density we first equilibrated the system for 5 million cycles, then gathering equilibrium statistics for just as many cycles. The pressure data obtained for $N=4000$ particles were plotted in the middle panel of Fig.\,5. In the same figure we also reported the outcome of the theory. We see that the data points are rather clearly located over four distinct pressure plateaus, each being representative of a different drop conformation of long life. While three of these plateaus have already been identified as representative of spherical, cylindrical, and slab-like drops, in the fourth plateau centered at $\rho=0.25$ a visual inspection of the system configuration reveals that the shape is novel: the liquid drop resembles either a {\em double cylinder} (DC, see Fig.\,6a) or a {\em punched slab} (PS, see Fig.\,6b). Owing to the periodicity of the simulation box, the DC and PS shapes would actually be similar: it suffices to move the DC center to a box vertex (and possibly rotate the structure just to improve visibility) to realize that a DC is in fact not dissimilar from a PS (Fig.\,7a). Similarly, by moving the hole center to a box vertex the PS of Fig.\,6b ends up looking like a DC (Fig.\,7b). However, while the boundary of the hole associated with a perfect DC is a square, the hole of the PS seen in Fig.\,6b is roughly circular; hence, it would be wrong to conclude that the two geometries are exactly the same up to a folding operation.

Let us first attempt to model the liquid drop as a perfect slab with a cylindrical hole (we call this geometric shape ``type-1 PS''). Let $d$ and $r<L_x/2$ be the slab thickness and hole radius, respectively. The interface area equals
\be
A=2(L_x^2-\pi r^2)+2\pi rd
\label{3-2}
\ee
with
\be
V=(L_x^2-\pi r^2)d=N_{\rm l}v_{\rm l}\equiv V_{\rm l}\,.
\label{3-3}
\ee
Upon eliminating $r$ in favor of $d$, we obtain:
\be
A(d)=\frac{2V_{\rm l}}{d}+2\sqrt{\pi(L_x^2d^2-V_{\rm l}d)}\,.
\label{3-4}
\ee
The previous equation is correct only provided $0<r<L_x/2$, or
\be
\frac{V_{\rm l}}{L_x^2}<d<\frac{4}{4-\pi}\frac{V_{\rm l}}{L_x^2}\,.
\label{3-5}
\ee
Upon taking $x=L_x^2d/V_{\rm l}-1=\pi r^2d/V_{\rm l}$ and $y=A/(2L_x^2)-1$, the problem is reduced to finding the absolute minimum of
\be
y=-\frac{x}{1+x}+\frac{V_{\rm l}}{L_x^3}\sqrt{\pi(x+x^2)}
\label{3-6}
\ee
in the interval $0<x<\pi/(4-\pi)$. The hole only forms if, besides the local minimum at $x=0$, a deeper (negative) $y$ minimum also occurs (at a certain $x_{\rm min}$ corresponding to $d=d_{\rm min}\propto N^{1/3}$). This requires $\rho$ to be less than a threshold density ($\simeq 0.175$, see Fig.\,8 below; beyond this density, the free energy becomes identical to $\Delta F_{\rm slab}$). With the $A$ so determined, the free energy of the PS is given by
\be
\Delta F=P_{\rm v}(v_{\rm v}-v)N+\gamma A\,.
\label{3-7}
\ee

However, representing the slab hole as a perfect cylinder is too rough an approximation, since the physical drop would certainly manage to avoid any sharp edges. A more realistic modelization (say, ``type-2 PS'') will entail a smooth and curved hole boundary, and the most natural solution would be the surface of the innermost half of a torus (IHT, namely the part of the torus which lies inside a cylinder having the same symmetry axis as the torus and radius equal to the torus major radius -- {\it i.e.}, to the distance $r$ from the center of the hole to the center of the tube). While the minor radius of the torus (that is, the radius of the tube section) has to be half of the slab thickness ($d/2$), the major radius must obey $d/2<r<L_x/2$ (when $r=d/2$ the hole closes and the torus becomes a horn torus).

We used Pappus' centroid theorem to compute the area and volume of the IHT of radii $R_<=d/2$ and $R_>=r$. From the general formulas,
\be
A_{\rm IHT}=2\pi^2R_<R_>-4\pi R_<^2\,\,\,\,\,\,{\rm and}\,\,\,\,\,\,V_{\rm IHT}=\pi^2R_<^2R_>-\frac{4}{3}\pi R_<^3\,,
\label{3-8}
\ee
it follows that the $A$ and $V$ in Eqs.\,(\ref{3-2}) and (\ref{3-3}) should be replaced with
\be
A=2(L_x^2-\pi r^2)+\pi^2dr-\pi d^2\,\,\,\,\,\,{\rm and}\,\,\,\,\,\,V=(L_x^2-\pi r^2)d+\frac{1}{4}\pi^2d^2r-\frac{\pi}{6}d^3\,.
\label{3-9}
\ee
Using the equation $V=V_{\rm l}$ to simplify the $A$ expression, we eventually obtain:
\be
A=\frac{2V_{\rm l}}{d}+\frac{1}{2}\pi^2dr-\frac{2}{3}\pi d^2\,,
\label{3-10}
\ee
where $r$ is the solution to
\be
r^2-\frac{1}{4}\pi dr+\frac{V_{\rm l}}{\pi d}+\frac{1}{6}d^2-\frac{L_x^2}{\pi}=0\,.
\label{3-11}
\ee
Should the solutions of Eq.\,(\ref{3-11}) be both valid, the one must be chosen that provides the minimum $A$ value. Then, the free energy follows from the general Eq.\,(\ref{3-7}).

Finally, we considered a drop having the shape of two equal cylinders crossing each other at right angles. We call this shape a DC (also the cylinder axes must intersect with one another). While the length of the axes is $L_x/2$, the cylinder radius $r$ should be consistent with the $V=V_{\rm l}$ condition. In order to compute the area and volume of a DC, it comes useful to consider the solid body which the two cylinders have in common, which goes under the name of {\em bicylinder} (or Steinmetz solid). Its area and volume can easily be obtained by multiple integration, and the results are $16r^2$ and $(16/3)r^3$. Clearly, the area and volume of a DC then read:
\be
A=4\pi rL_x-16r^2\,\,\,\,\,\,{\rm and}\,\,\,\,\,\,V=2\pi r^2L_x-\frac{16}{3}r^3\,.
\label{3-12}
\ee
From $V=V_{\rm l}$ a third-order algebraic equation is obtained for $r$, which turned out to have only one solution satisfying $0<r<L_x/2$. Upon plugging this $r$ in $A$, we again derive $\Delta F$ from Eq.\,(\ref{3-7}).

For $T=0.72$, the free-energy curves for the three just considered shapes were plotted as a function of the density in the top panel of Fig.\,8. In the same picture, the lower envelope of the free energies for the common shapes was reported for comparison (red line). We see that none of the PS nor the DC ever provide the shape of the stable drop. However, a type-2 PS with $r=L_x/2$ is nearly stable close to $\rho_{\rm cyl-slab}\simeq 0.25$, which may explain why this drop shape was observed in our simulations from scratch (as seen in Fig.\,8, suitable type-1 PS and DC also exist that offer not too bad solutions near $\rho_{\rm cyl-slab}$). Beyond $\rho=0.24$, the optimal type-2 PS has $r=d/2$, meaning that the hole eventually closed and the torus became a horn torus (however, this degenerate type-2 PS is obviously no longer reliable as a drop shape). Finally, notice that changing $T=0.72$ to $T=0.90$ does not modify the theoretical picture in any respect. The conclusion is that, according to the present theory, the three novel shapes considered in this section are metastable. Hence, the sequence of stable drops for $T=0.72$ remains the same identified in Sect.\,III-A for $T=0.90$ and $a=1$.

\section{Conclusions}

Below the critical temperature, the structure of a heterogeneous fluid simulated under periodic boundary conditions is sensitive to the imposed density. Each change of conformation observed within the binodal line goes along with a characteristic fall in the pressure. In this paper, the by-now classical evidence of spherical, cylindrical, and slab-like liquid drops was re-examined from the viewpoint of a theory having its roots in a proof-of-concept calculation found in Ref.~\cite{MacDowell1}. In spite of its simplicity, this theory well reproduced the behavior of a Lennard-Jones fluid along the liquid-vapor region, for both a cubic and an elongated box.

Further drop shapes emerged near the crossover region from cylindrical to slab-like when simulations, rather than being performed in a sequence as is common practice, were started for each density from a face-centered-cubic configuration. In this case, the liquid drop occasionally exhibited the shape of a punched slab with a roughly circular hole. Employing our theory, we found that this peculiar shape of drop is actually only metastable, {\it i.e.}, it is a long-lived conformation which however is doomed to decay.

\newpage
%
%
\begin{figure}
\begin{center}
\includegraphics[width=12.0cm]{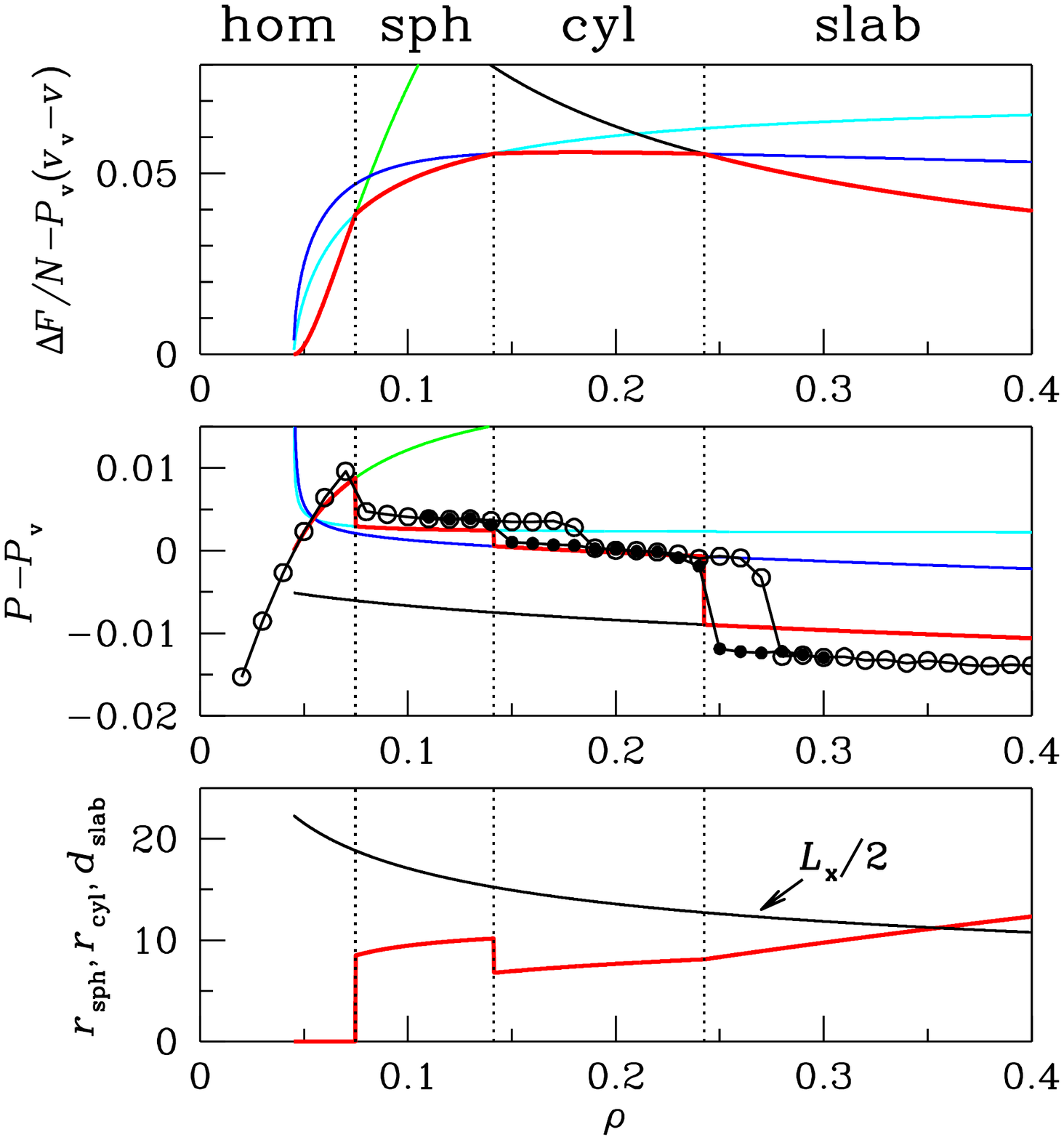}
\caption{(Color online) Comparison between theory and MC data\,\cite{Abramo2} from sequential simulations of the cut-and-shifted LJ model in a periodic cubic box ($N=4000$ and $T=0.90$). In order to equilibrate the system $10^6$ MC cycles were produced at each state point, followed by other $4\times 10^6$ cycles over which equilibrium averages were computed. Top: $\Delta F/N-P_{\rm v}(v_{\rm v}-v)$ for the various competing ``phases'' (green, ``hom''; cyan, ``sph''; blue, ``cyl''; black, ``slab''). The lower envelope of all the free-energy curves (thick red line) marks the equilibrium curve. The vertical lines mark the location of the shape transitions (see Eqs.\,(\ref{2-12})). Middle: system pressure relative to $P_{\rm v}$. The dots are the MC data from Ref.\,\cite{Abramo2} (open dots, forward trajectory; full dots, backward trajectory) while the thick red line is the equilibrium pressure according to our theory. Bottom: the equilibrium drop size (thick red line, see Eqs.\,(\ref{2-13})) vs. half box size (black line).}
\label{fig1}
\end{center}
\end{figure}

%
%
\begin{figure}
\begin{center}
\includegraphics[width=12.0cm]{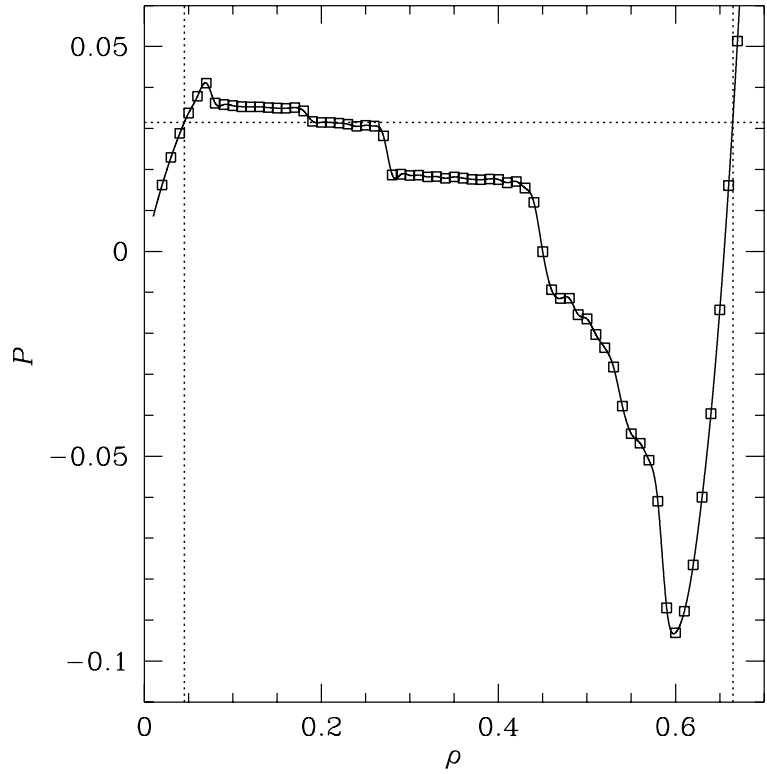}
\caption{MC pressure data for the cut-and-shifted LJ model in a periodic cubic box with $N=4000$ and $T=0.90$ (only the forward trajectory is shown). The horizontal line lies at the level of $P_{\rm v}$. The vertical lines mark the liquid-vapor coexistence densities, $\rho_{\rm v}$ and $\rho_{\rm l}$.}
\label{fig2}
\end{center}
\end{figure}

%
%
\begin{figure}
\begin{center}
\includegraphics[width=12.0cm]{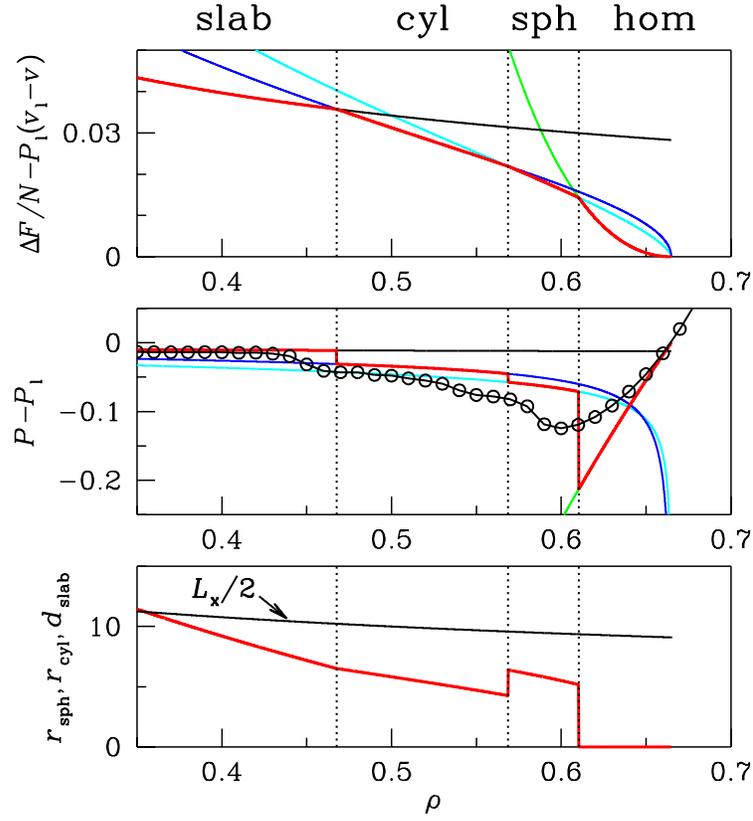}
\caption{(Color online) Comparison between theory and MC data from sequential simulations of the cut-and-shifted LJ model in a periodic cubic box ($N=4000$ and $T=0.90$). Same as Fig.\,1 but the high-density region is shown here.}
\label{fig3}
\end{center}
\end{figure}

%
%
\begin{figure}
\begin{center}
\includegraphics[width=12.0cm]{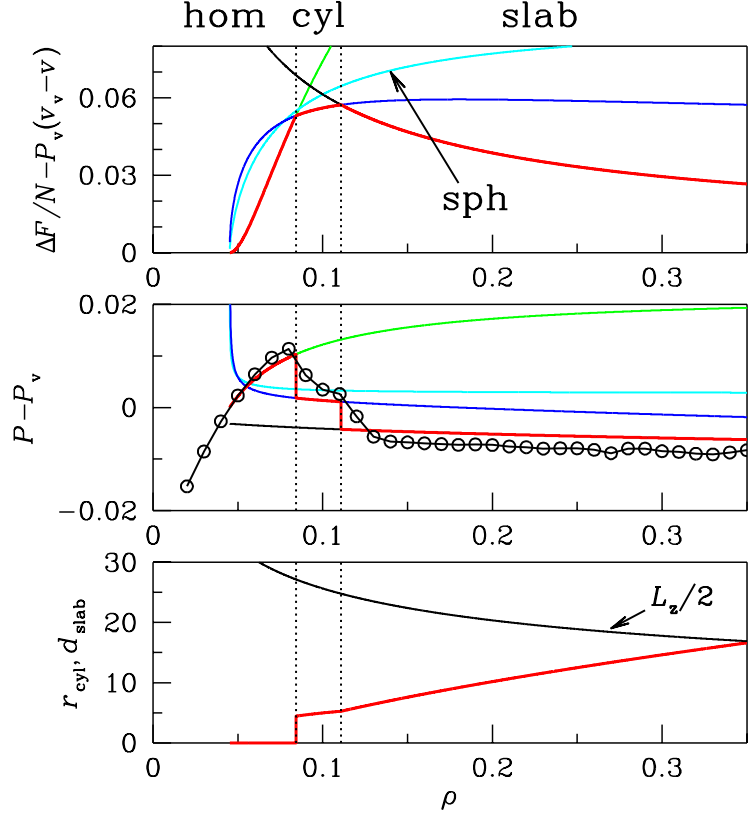}
\caption{(Color online) Comparison between theory for $a=3$ and MC data from sequential simulations of the cut-and-shifted LJ model in a periodic cuboidal box ($N=1500,T=0.90$, and $\beta\gamma=0.175$). In strike contrast with the $a=1$ case, the spherical drop would never be stable for $a=3$. The cylindrical drop is only observed in a narrow range of densities close to 0.1 whereas the slab-like drop is stable over a much wider density interval.}
\label{fig4}
\end{center}
\end{figure}

%
%
\begin{figure}
\begin{center}
\includegraphics[width=12.0cm]{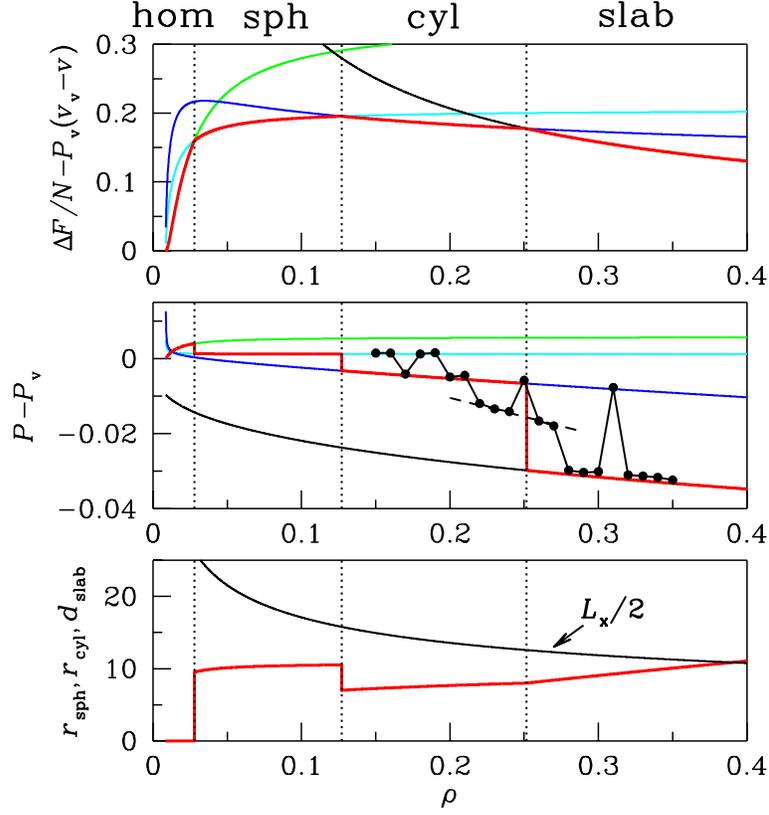}
\caption{(Color online) MC pressure data for the cut-and-shifted LJ model in a periodic cubic box ($N=4000$ and $T=0.72$) vs. theory (for this case, $P_{\rm v}=0.0062,\rho_{\rm v}=0.0086,\rho_{\rm l}=0.7722,\beta\gamma=0.78$~\cite{Trokhymchuk}, whereas the value of $K_{\rm v}$, 173.62, resulted from a long $NPT$ simulation of the liquid at $P_{\rm v}$). The data points (full dots in the middle panel) were obtained independently from one another, by starting the simulation from scratch at each density. We see that each point lies on a theoretical line, except for the points around $\rho=0.25$ which appear to lie on a different plateau (dashed line). In the corresponding range of densities the liquid drop resembles either a DC or a PS.}
\label{fig5}
\end{center}
\end{figure}

%
%
\begin{figure}
\begin{center}
\begin{tabular}{cc}
\includegraphics[width=10.0cm]{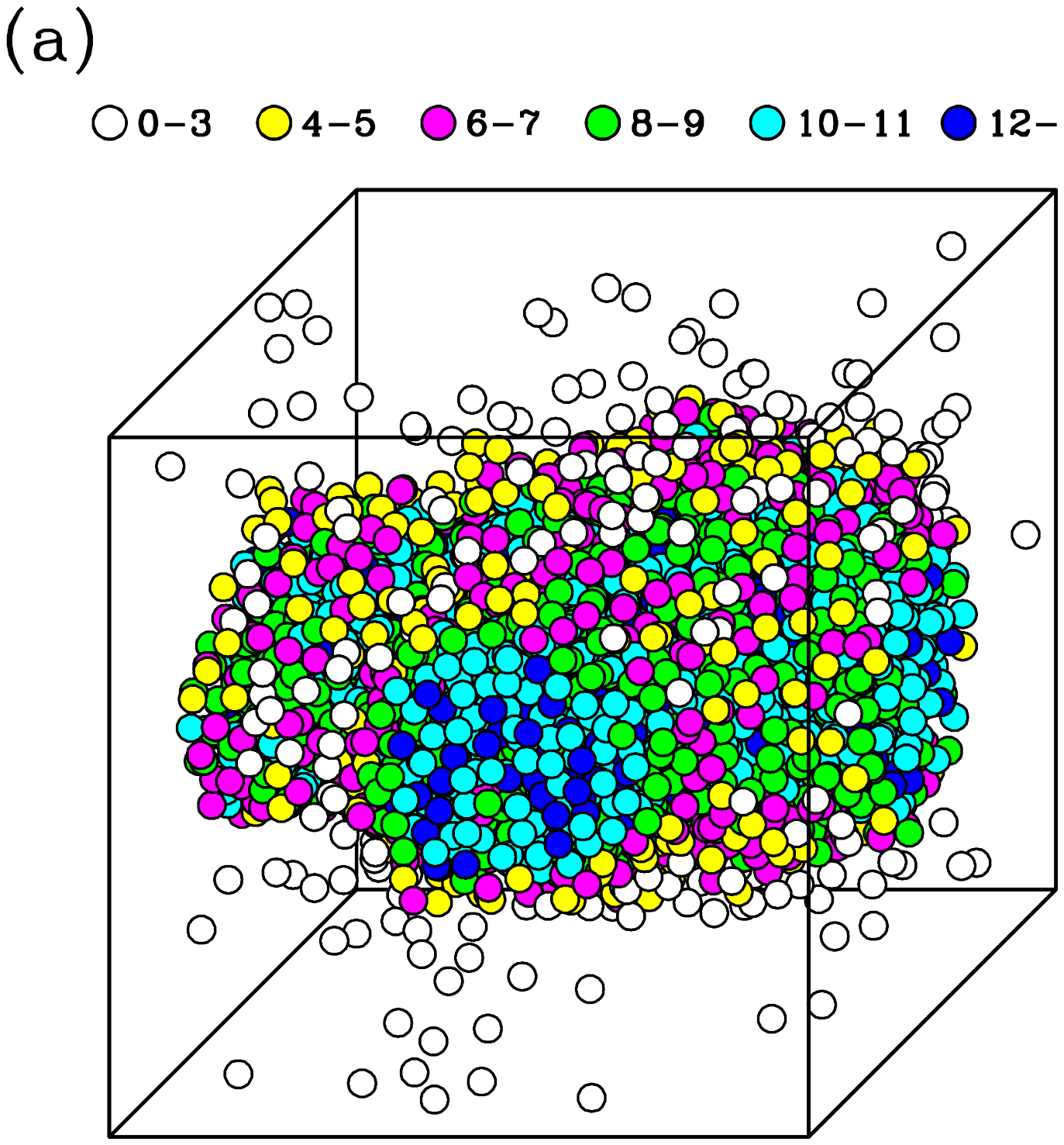} &
\includegraphics[width=10.0cm]{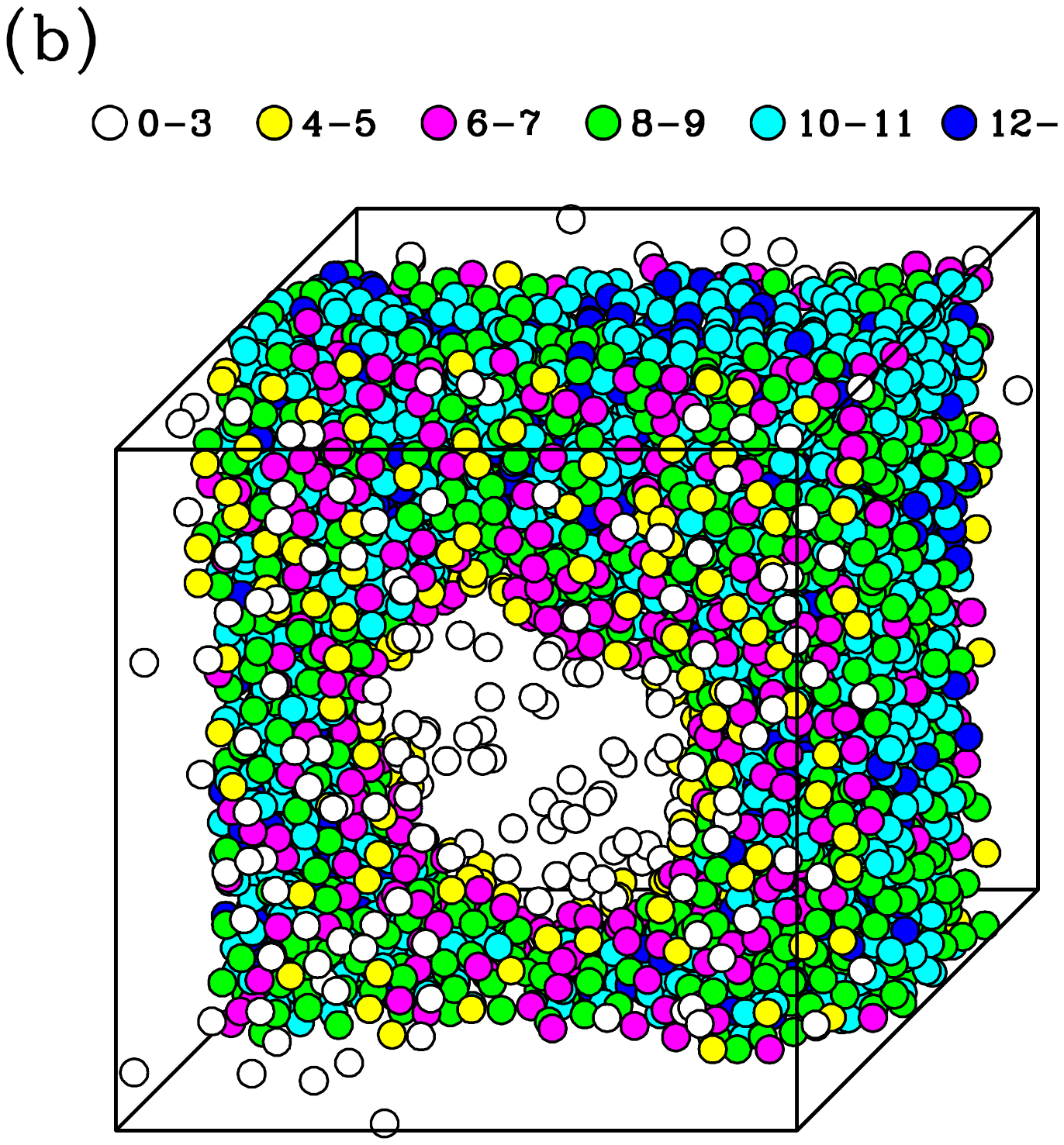}
\end{tabular}
\caption{(Color online) Cut-and-shifted LJ model for $N=4000$ and $T=0.72$: results from MC simulations started from scratch. (a) A system snapshot taken at the density $\rho=0.24$; (b) another snapshot taken at $\rho=0.26$. In these pictures, each particle was given an effective diameter $\sigma$. The symbol colors were chosen according to the number of nearest neighbors each particle has, in turn defined as the number of particles within a distance of $1.45\sigma$ from the given particle ($1.45\sigma$ being the position of the first non-zero minimum of the radial distribution function of the liquid at $\rho_{\rm l}$). While the liquid drop in (a) resembles a DC, the drop in (b) looks like a PS.}
\label{fig6}
\end{center}
\end{figure}

%
%
\begin{figure}
\begin{center}
\begin{tabular}{cc}
\includegraphics[width=10.0cm]{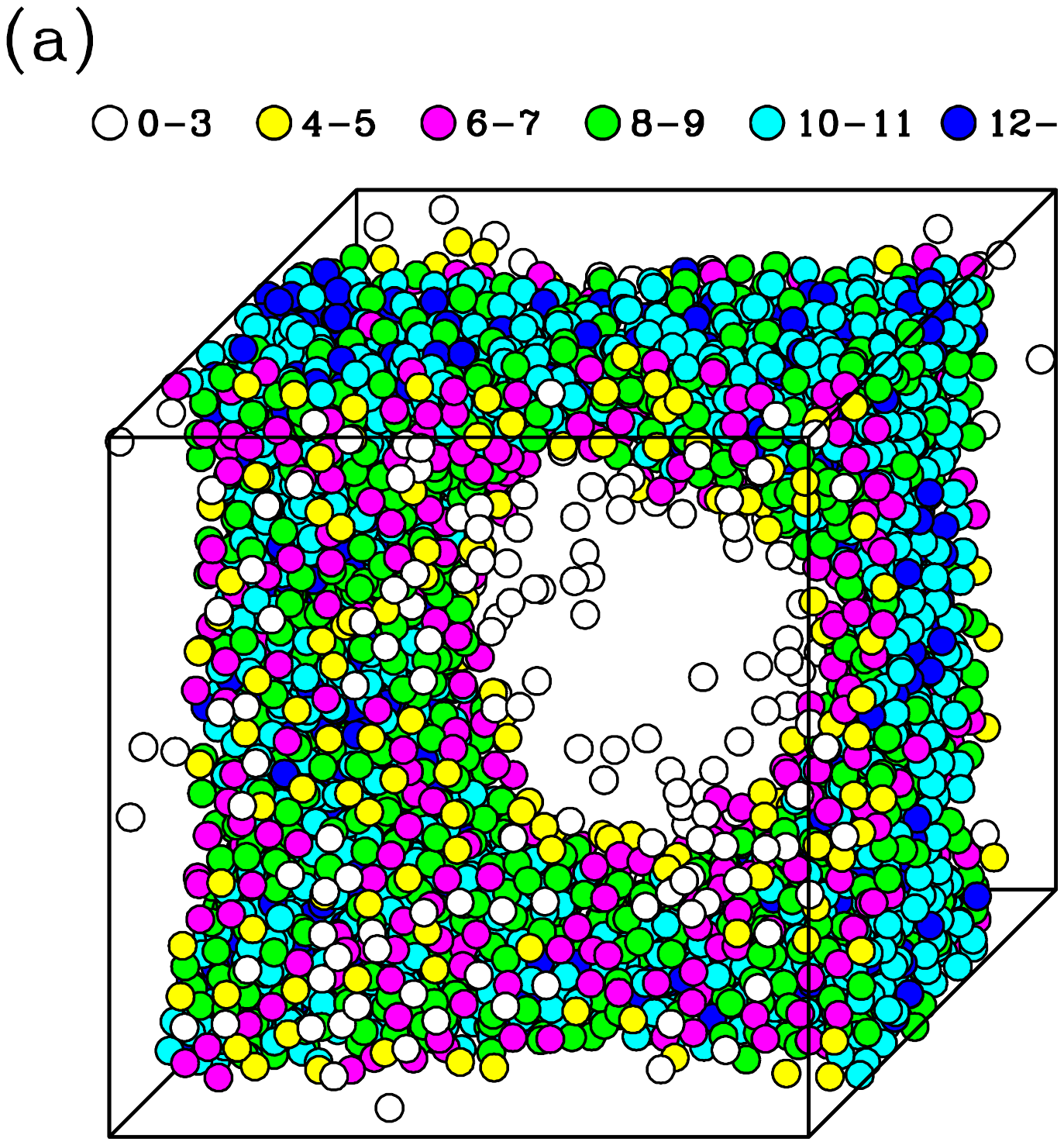} &
\includegraphics[width=10.0cm]{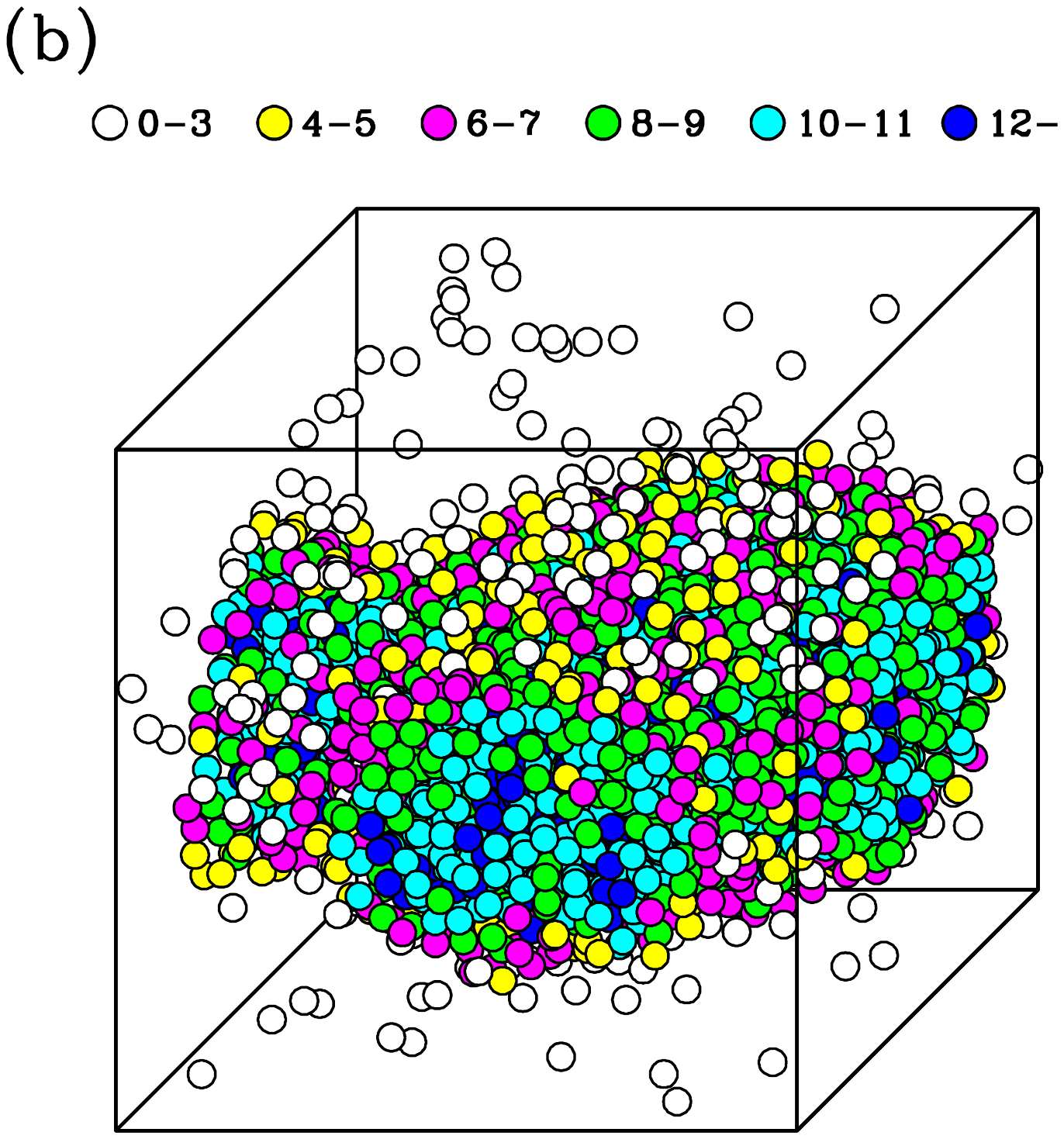}
\end{tabular}
\caption{(Color online) Same as in Fig.\,6, but after the particle coordinates have been transformed as explained in the main text. The outcome was that a DC became a PS, and vice versa.}
\label{fig7}
\end{center}
\end{figure}

%
%
\begin{figure}
\begin{center}
\includegraphics[width=12cm]{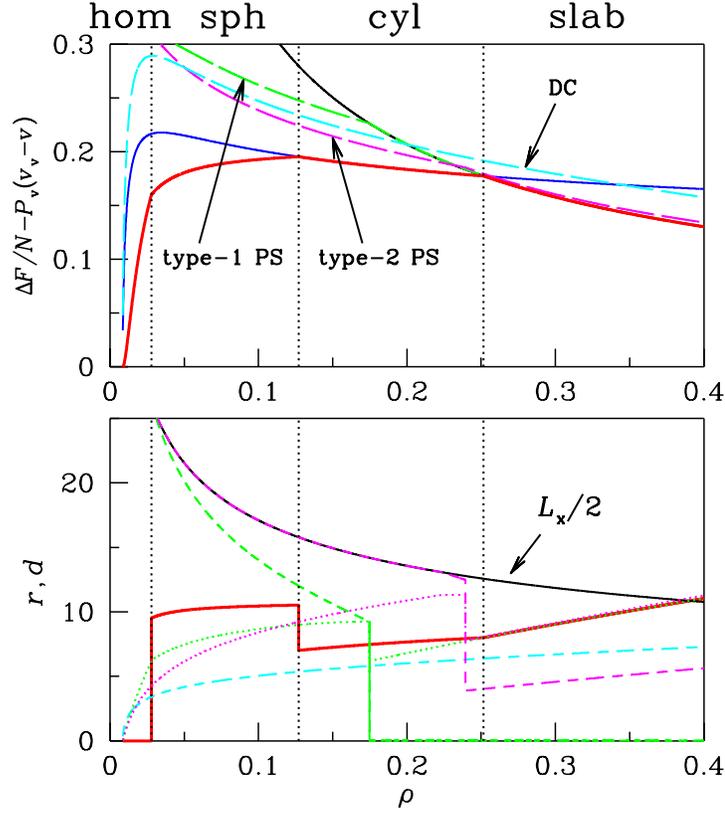}
\caption{(Color online) Shape transitions for the cut-and-shifted LJ model in a periodic cubic box ($N=4000$ and $T=0.72$): theoretical results (see parameters in Fig.\,5 caption). Top: $\Delta F/N-P_{\rm v}(v_{\rm v}-v)$ for various competing ``phases'' (blue full line, ``cyl''; black full line, ``slab''; green long-dashed line, type-1 PS; cyan long-dashed line, DC; magenta long-dashed line, type-2 PS -- notice that the free-energy curves for ``hom'' and ``sph'' are not shown). The thick red line marks the equilibrium curve (that is, the lower envelope of all the free-energy curves, including ``hom'' and ``sph''). The vertical lines mark the location of the shape transitions. Bottom: the equilibrium drop size (thick red line) vs. half box size (black full line). The dotted and dashed lines respectively refer to the optimal $d$ and $r$ values for the two types of PS and the DC.}
\label{fig8}
\end{center}
\end{figure}

\end{document}